
%
%
%
%
%
%
\input harvmac.tex
\lref\GM{D. Gross. and A. Migdal, Phys. Rev. Lett. {\bf 64} (1990)
127.}
\lref\pasq{I. Klebanov and A. Pasquinucci, Princeton preprint
PUPT-1313.}
\lref\DS{M. Douglas and S. Shenker, Nucl. Phys. {\bf B335} (1990) 635.}
\lref\BK{E. Brezin and V. Kazakov, Phys. Lett. {\bf 236B} (1990) 144.}
\lref\TP{E. Witten, Nucl. Phys. {\bf B340} (1990) 281.}
\lref\DW{R. Dijkgraaf and E. Witten, Nucl. Phys. {\bf B342} (1990) 486.}
\lref\LZ{B. Lian and G. Zuckerman, Phys. Lett. {\bf 254B} (1991) 417.}
\lref\Lz{B. Lian and G. Zuckerman, Comm. Math. Phys. {\bf 135} (1991)
547.}
\lref\BMP{P. Bouwknegt, J. McCarthy, and K. Pilch, CERN preprint
CERN-TH.6162/91, to appear in Comm. Math. Phys.}
\lref\Pol{A. Polyakov, Mod. Phys. Lett. {\bf A6} (1991) 635.}
\lref\GR{E. Witten, IAS preprint IASSNS-HEP 91/51.}
\lref\KP{I. Klebanov and A. Polyakov, Mod. Phys. Lett. {\bf A6} (1991)
3273.}
\lref\kut{M. Bershadsky and D. Kutasov, Harvard and Princeton preprint
HUPT-92/A016, PUPT-1315.}
\lref\K{I. Klebanov, Mod.
Phys. Lett. {\bf A7} (1992) 723.}
\lref\Zwi{E. Witten and B. Zwiebach, IAS preprint IASSNS-HEP 92/04.}
\lref\KMS{D. Kutasov, E. Martinec, and N. Seiberg,
Phys. Lett. {\bf 276B} (1992) 437.}
\lref\Kac{S. Kachru, Princeton preprint PUPT-1305, to appear in Mod.
Phys. Lett. {\bf A}.}
\lref\Nla{H.S. La and P. Nelson, Nucl. Phys. {\bf B332} (1990) 83.}
\lref\alv{L. Alvarez-Gaume, C. Gomez, G. Moore, and C. Vafa, Nucl. Phys.
{\bf B303} (1988) 455.}
\lref\dn{J. Distler and P. Nelson, Comm. Math. Phys. {\bf 138} (1991)
273.}
\lref\Doyle{M. Doyle, Princeton preprint PUPT-1296.}
\lref\Nelson{P. Nelson, Phys. Rev. Lett. {\bf 62} (1989) 993.}
\lref\Li{M. Li, University of California at Santa Barbara preprint
UCSBTH-91-47.}
\lref\CDK{N. Chair, V.K. Dobrev, and H. Kanno, ICTP preprint IC/92/17.}
\lref\Dot{V. Dotsenko, LPTHE preprint PAR-LPTHE 92-4.}
\lref\Douglas{M. Douglas, Phys. Lett. {\bf 238B} (1990) 176.}
\lref\DVV{R. Dijkgraaf, E. Verlinde and H. Verlinde, Nucl. Phys. {\bf
B348} (1991) 435.}
\lref\Ulf{U.H. Danielsson and D.J. Gross, Nucl. Phys. {\bf B366} (1991)
3.}
\lref\Mike{D.J. Gross, I.R. Klebanov, and M.J. Newman, Nucl. Phys.
{\bf B350} (1990) 621.}
\lref\moore{G. Moore, Rutgers preprint RU-91-12.}
\lref\imbimbo{C. Imbimbo, S. Mahapatra and S. Mukhi, Tata Institute
preprint TIFT/TH/91-27.}
\lref\govind{S. Govindarajan, T. Jayaraman, V. John, and P. Majumdar,
Institute of Mathematical Sciences preprint IMSc-91/40.}
\Title{PUPT-1314}
{\vbox{\centerline{Extra States and Symmetries}
   \vskip .1in\centerline{in}
   \vskip .1in \centerline {$D<2$ Closed String Theory}}}
\centerline{{Shamit Kachru}\footnote{$^\dagger$}
{Supported in part by an NSF Graduate Fellowship}}
\bigskip\centerline{\it Joseph Henry Laboratories}
\centerline{\it Princeton University}\centerline{\it Princeton, New Jersey
08544}
\centerline {\it kachru@puhep1.princeton.edu}
\vskip .3in
We show that there is $(p-1)(p^\prime -1)$ dimensional semi-relative BRST
cohomology
at each non-positive ghost number in
the ($p,p^\prime$) minimal conformal field
theory coupled to two dimensional quantum gravity.  These closed string
states are related
to currents and symmetry charges of `exotic' ghost number.
We investigate the symmetry structure generated by the most conventional
currents (those of vanishing total ghost number), and make a conjecture about
the extended algebra which results from incorporating the currents at
negative ghost numbers.

\Date{April 1992} 
\newsec{Introduction}

Our understanding of $D \le 2$ string theories has been greatly
improved in the past two years by the realization that the double-scaling
limit of matrix models can be used to probe these string theories to all
orders in perturbation theory (and sometimes beyond)\GM \DS \BK, and by the
concurrent realization that these models are closely related to
topological field theories \TP \DW.  However, we still lack a complete
understanding of how one can obtain the remarkable results of matrix
models working directly in a continuum (Liouville theory $\times$ conformal
matter)
formulation.

A focus of recent studies has been the role of the `extra' physical states
called special states,
first discovered and studied in the $c=1$ matrix model in
\Mike \Ulf \moore~ and later uncovered by Polyakov in the continuum version
of this theory~\Pol.  These states (and others, which we will also call
special states) have since been found rigorously in
the BRST cohomology of the continuum theory \Lz\BMP.
They play a crucial role in our
understanding of the correspondence between the continuum theory and the
$c=1$ matrix model.  In
particular, some of the special states are responsible for generating the
$W_{\infty}$
symmetry of the theory \GR \KP. In addition,  one can see the analogue of the
free
fermion structure of the matrix models by analyzing the `ground ring' formed by
the operators of
vanishing ghost number (the ghost number increases by one in going from
states to operators; in this paper we use the convention that the
SL(2,$\cal C$) invariant vacuum has ghost number -1).
The $W_{\infty}$ symmetry leads to
powerful non-linear Ward Identities which can be used to greatly
simplify computations in this theory \K \Zwi \pasq, and the fact that the
tachyon states of the theory form modules of the ground ring
also leads to certain identities the correlation functions must satisfy
\KMS \Kac \kut.

While the special states have proven to be of great utility in the $D =
2$ theory, their presence in the $D < 2$ theories \LZ \BMP~ (minimal models
coupled to 2D quantum gravity) has not yet been fully understood and
exploited.  In this paper, we will show that there are infinitely more `extra'
physical states than those which have been widely discussed to date, in the
{\it {semi-relative cohomology}} of the $D < 2$ theories (a similar
observation has been made regarding the $D = 2$
theory by Witten and Zwiebach \Zwi).
Some of these states correspond to standard currents (of total ghost number
zero{\footnote{$^1$}{Actually, since as we will see these currents have
non-vanishing chiral ghost numbers which sum to zero, they aren't quite
like conventional currents}}), which generate
symmetries of the theories.  In this paper we determine the symmetry
algebra of the ($p,p^\prime$) models (by which we mean 2-d gravity coupled to
($p,p^\prime$) minimal matter, with $p>p^\prime$ and $p,p^\prime$
relatively prime).  We also provide a preliminary exploration of the
action of the associated conserved charges on the physical states.  We find
that the charges of total ghost number zero  act trivially on the regular
tachyon states (dressed minimal
CFT primary
fields) of these models, but expect that the full algebra of exotic
symmetries will act non-trivially on
the infinite tower of states discovered in \LZ.
\newsec{Formalism}
\subsec{Physical State Conditions in (Bosonic) String Theory}
Since the new string states we find are in the $\it {semi-relative}$
cohomology, and satisfy slightly weaker conditions than the `normal'
physical state conditions one is used to from critical string theory, we
take a moment to review here how this set of conditions arises in the
operator formalism.  We will follow
the treatment of Distler and Nelson \dn. For more details on the
operator formalism see also \alv \Nelson \Nla \Doyle.

Denote by $\cal M_{\rm g,n}$ the space of conformally distinct smooth
Riemann surfaces with g handles and n marked points, and by $\cal P_{\rm
g,n}$ the space of such surfaces where in addition we have chosen a local
coordinate $z_{i}$ centered around each of the marked points.  Then a
conformal field theory associates to each element $X \in \cal P_{\rm g,1}$ a
vector $\langle X \vert$ in a Hilbert space $\bar {\cal H}$.
One can
think of obtaining this state by performing a functional integral over all of
the Riemann surface
except for an excised disk around the marked point, yielding a wavefunctional
of the
boundary conditions one chooses to insert on the disk: If one inserts
boundary conditions corresponding to the state $\vert \psi \rangle \in
\cal H$, then the resulting complex number is $\langle X \vert
\psi\rangle$.
Similarly, to an element of $P_{\rm g,N}$ a conformal field theory
associates a vector in the N-fold tensor product ${\bar {\cal H}}^{\otimes \rm
N}$ (actually, more precisely one gets a vector in ${\bar {\cal
H}}^{\otimes \rm Q}\otimes {\cal H}^{\otimes N-Q}$ where $Q \le N$ is
determined by the orientation chosen for the boundaries of the disks
surrounding the punctures; this needn't concern us here).

Given any N vectors $\vert \psi_{\rm 1}\rangle,...,\vert \psi_{\rm
N}\rangle~\in \cal H$,
we wish to construct a differential form of degree $3g-3+N$ on $\cal
M_{\rm g,N}$,  which we can then integrate over $\cal M_{\rm g,N}$ to give us
the genus g N-point function of the operators $\psi_{\rm
1},...,\psi_{\rm N}$ associated with the vectors.
First, we will construct such a form on $\cal P_{\rm
g,N}$, by specifying its action on
any $3g-3+N$ tangent vectors to $\cal P_{\rm g,N}$ at some point $Y$.
Take an N-tuple of vector fields $\vec v = (v_{1},..,v_{N})$ (one $v_{\rm
i}$ in the
neighborhood of each of the punctures) which do
not belong to the `Borel subalgebra' $B(Y,z_{1},...,z_{n})$, where the
$z_{i}$ are local coordinates vanishing at the punctures
$P_{i}$.{\footnote{$^2$}{Elements of $B$ are sets of $\vec v_{i}$ which are all
the restriction of a single vector
field $\vec v$ holomorphic on $Y ~{\backslash \ } \{P_{1},...,P_{n}\}.$}}
These together correspond to
a nonzero tangent vector $V$ to $\cal P_{\rm g,N}$ at $Y$.  Define:
$$b[\vec v] = \sum_{i=1}^N \oint b_{zz}^{(i)}(z)v_{i}^{z}(z) dz
\eqno (2.1)$$
where $b_{zz}^{(i)}$ is the spin-2 anti-commuting ghost field of the $bc$
system
introduced in string theory to fix diffeomorphism invariance, and acts on the
i-th copy of $\cal H$.
Then we define the
differential form associated to $\vert \psi_{\rm 1}\rangle \otimes ... \otimes
\vert \psi_{\rm N}\rangle$ by:
$$\tilde \Phi (V_{\rm 1},...,V_{\rm 3g-3+N},\bar V_{\rm 1},...,\bar V_{\rm
3g-3+N})
\equiv$$
$$ \langle Y \vert b[\vec v_{\rm 1}]...b[\vec v_{\rm 3g-3+N}]\bar
b[{{\bar {\vec v}}_{\rm 1}}]...\bar b[{\bar {\vec v}}_{\rm 3g-3+N}]\vert
\psi_{\rm 1}\rangle\otimes ...\otimes \vert \psi_{\rm N}\rangle\eqno
(2.2)$$
Here, $\bar b$ is the anti-holomorphic counterpart to $b$, and the $\bar
{\vec v}$~s correspond to the conjugates $\bar V_{i}$ to the $V_{i}$.

Now, (2.2) does give us a form on $\cal P_{\rm g,N}$; what we really
want is a form on $\cal M_{\rm g,N}$.  The simplest possibility is that
$\tilde \Phi$ is simply the $\it lift$ of the desired form $\Phi$ on
$\cal M_{\rm g,N}$:
$$\tilde \Phi = \pi^{*}\Phi~\eqno (2.3)$$
where $\pi : \cal P_{\rm g,N} \rightarrow \cal M_{\rm g,N}$ is the
projection map for the $\cal P_{\rm g,N}$ bundle over $\cal M_{\rm
g,N}$.
This is the case exactly when each of the states satisfies:
$$L_{n}\vert \psi \rangle = b_{n}\vert \psi \rangle = \bar L_{n}\vert
\psi\rangle = \bar b_{n}\vert \psi \rangle~=~0,~~~~n\ge 0\eqno
(2.4)$$
where the $b_{n}$ and the $L_{n}$ are defined in the normal way as the Laurent
coefficients of the
$b$ ghost field and the holomorphic stress-energy tensor, and likewise for
their
counterparts $\bar b_{n}$ and $\bar L_{n}$.
These are the most familiar physical state conditions from critical
string theory, termed in \Nelson ~the `Strong Physical State
Conditions.'  We can get weaker physical state conditions by
realizing that (2.3) is not really necessary for us to reconstruct a
form $\Phi$ on $\cal M_{\rm g,N}$ from $\tilde \Phi$.

Choose a section $\sigma: \cal M_{\rm g,N} \rightarrow \cal P_{\rm
g,N}$.  Then we can imagine pulling back $\tilde \Phi$ using this section:
$$\Phi = \sigma ^{*}\tilde \Phi .\eqno (2.5)$$
While the resultant $\Phi$ obviously depends on the section $\sigma$ chosen,
it turns out that if $\tilde \Phi$ is a $\it closed$ differential form,
then $\Phi$ in (2.5) yields a well-defined cohomology class irrespective
of which section we choose.  Hence, its integral (modulo fixing
conditions at the boundaries of $\cal M_{\rm g,N}$) will be well
defined.

In fact, $\tilde \Phi$ is indeed a closed form whenever the states
involved in its definition in (2.2) obey
a $\it weaker$ condition than (2.4), namely that
$$(Q + \bar Q)\vert \psi \rangle = \rm 0\eqno (2.6)$$
where $Q$ and $\bar Q$ are the holomorphic and anti-holomorphic BRST
operators.

However, another complication arises: There is in general an obstruction to
choosing a global section
$\sigma$ (the bundle $\pi : \cal P_{\rm g,N} \rightarrow \cal M_{\rm g,N
}$
is nontrivial).  What we $\it can$ do is find sections which are well defined
up to constant phase jumps across coordinate patch overlaps.  If we can
construct $\tilde \Phi$ so that it is insensitive to such jumps, then
again $\Phi$ defined via (2.5) will be acceptable.  The necessary
conditions for this to occur are:
$$(Q+\bar Q)\vert \psi \rangle = (L_{\rm 0} - \bar L_{\rm 0}) \vert \psi
\rangle = (b_{\rm 0} - \bar b_{\rm 0})\vert \psi \rangle = 0.\eqno
(2.7)$$

Actually, we will only consider states which are annihilated by both $Q$
and $\bar Q$ and which are in the
nullspaces of both $L_{\rm 0}$ and $\bar L_{\rm 0}$.  With this
proviso, the conditions (2.4) correspond to states of the closed string which
are in the chiral $\it relative$
BRST cohomology (they are annihilated by $b_{\rm 0}$ and $\bar b_{\rm 0
}$):
The chiral halves would be ${\it bona~ fide}$
open string states.  The conditions (2.7) correspond to states of the
closed string which are in the $\it semi-relative$ cohomology.

\subsec{Symmetries and Descent Equations}
Let us briefly recall the formalism for currents conserved up to BRST
commutators, following \Zwi .  We assume we are working in the context
of a two dimensional quantum field theory, in local complex coordinates
on some Riemann surface.  A strictly conserved current
corresponds to a one form
$$\Omega ^{\rm 1} = J_{z}dz - J_{\bar z} d\bar z \eqno (2.8)$$
which is closed,
$$d\Omega ^{\rm 1} = 0. \eqno (2.9) $$
 One then obtains a
conserved charge  by integrating this one form around a contour,
the charge being conserved in the sense that this integral gives the
same result for homologous contours [we will throughout use the
convention that ${1\over 2\pi i}\oint {dz\over z} = -{1\over 2\pi i}\oint
{d{\bar z}\over {\bar z}} = 1$].

However, in BRST cohomology $Q_{TOT}=Q+\bar Q$ commutators are trivial, so in
BRST
quantization a well-defined charge on the Hilbert space of physical
states need only be invariant up to $Q_{TOT}$ commutators when evaluated on
homologous cycles.  This corresponds to the case that $\Omega ^{\rm 1}$
is only closed up to a $Q_{TOT}$ commutator of some two form $\Omega ^{\rm 2}$:
$$ d\Omega ^{\rm 1} = \{Q_{TOT}, \Omega ^{\rm 2}\}. \eqno (2.10)$$
As discussed in \Zwi, when $\Omega ^{\rm 2} \not= 0$, the Ward
Identity corresponding to the symmetry generated by this current can
be nonlinear.

Of course, in order for the conserved charge to map physical states to
physical states, it must commute with the BRST operator $Q_{TOT}$.  This occurs
if
there is a zero form $\Omega^{\rm 0}$ such that:
$$ d\Omega^{\rm 0} = \{Q_{TOT}, \Omega^{\rm 1}\}. \eqno (2.11)$$
It is then the case that $\{Q_{TOT}, \Omega^{\rm 0}\} = 0$ automatically.

Thus, the whole hierarchy of descent equations is given by:
$$ 0 = \{Q_{TOT}, \Omega ^{\rm 0}\}$$
$$ d\Omega ^{\rm 0} = \{Q_{TOT}, \Omega ^{\rm 1}\}\eqno (2.12)$$
$$ d\Omega ^{\rm 1} = \{Q_{TOT}, \Omega ^{\rm 2}\}$$
These equations imply that $\Omega ^{\rm 1}$ is a conserved current in
the BRST formalism.  So in particular, finding symmetries of the theory
is equivalent to finding zero forms $\Omega ^{\rm 0}$ which give non-trivial
results for $\Omega ^{\rm 1}$ when plugged into the descent equations.

The descent equations can be re-written in terms of states instead of
operators, with the results that for a given $\vert \Omega ^{\rm 0}\rangle$
$$\vert \Omega ^{\rm 1}_{z}\rangle = b_{-1}\vert \Omega ^{\rm 0}\rangle$$
$$\vert \Omega ^{\rm 1}_{\bar z}\rangle = \bar b_{-1}\vert \Omega ^{\rm
0}\rangle \eqno (2.13)$$
$$\vert \Omega ^{\rm 2}\rangle = b_{-1}\bar b_{-1} \vert \Omega ^{\rm
0}\rangle $$
Hence, to find conventional symmetries of the closed string theory,
which correspond to charges which have ghost number zero, one
must find BRST invariant states $\vert \Omega ^{\rm 0}\rangle$~in the
semi-relative cohomology at ghost number 0. Then the corresponding operator
$\Omega
^{\rm 1}$ obtained from the descent equations will also have vanishing
ghost number, as will the charge obtained from its contour integral.

\newsec{Extra States in Semi-relative Cohomology}
\subsec{Proof of Existence of Extra States}
In order to clarify the discussion below, we introduce some notation.
We will denote by $\rm H_{rel}^{n}(k)$ the chiral relative BRST cohomology of
the
($p,p^\prime$) minimal model coupled to Liouville theory at ghost number n
built on the Liouville Fock space with momentum k.
$\rm H_{abs}^{n}(k)$ will denote the corresponding absolute BRST
cohomology.

One can find the analogue  of the ground ring in the $D<2$ string
theories by examining the appropriate cohomology of the ($p,p^\prime$) minimal
model coupled
to quantum gravity.  This has been done by Kutasov, Martinec, and
Seiberg \KMS, who find that in the ($p,p^\prime$) model the ground ring
consists of
operators $\sl R_{\rm n,n'}$ with $1\le n \le p-1$ and $1 \le n^\prime \le
p^\prime-1$.  This
ring is the ring generated by $\sl X = R_{\rm 2,1}$ and $\sl Y = R_{\rm
1,2}$:
$$ \sl R_{\rm n,n'} = \sl X^{\rm n-1}\sl Y^{\rm n'-1}\eqno (3.1)$$
(the relations $\sl X^{\rm p-1} = \sl Y^{\rm p'-1} = \rm 0$ hold).

Let us denote by $\sl O_{\rm n,n'}$ the holomorphic chiral half of $\sl
R_{\rm n,n'}$.  If we call the Liouville momentum of the
operator $\sl O_{\rm n,n'}$ $p_{n,n'}$,
then we see that in the notation introduced above, these ring elements
correspond to
nontrivial cohomology classes $\vert \sl O_{\rm n,n'}\rangle ~\in ~\rm
H_{rel}^{(-1)}(p_{n,n'})$ in the chiral BRST cohomology.  In fact, these are
the
only such classes in the relative cohomology at their respective momenta
and ghost number -1.
Now, recall the result that:
$$ \rm H_{abs}^{(n)}(p) \simeq H_{rel}^{(n)}(p) \oplus c_{0}H_{rel}^{(n-1)}(p)
\eqno
(3.2)$$
Since we know that $\rm
H_{rel}^{(0)}(p_{n,n'}) = \emptyset$ for all n,n$^\prime$, it follows that
$$ \rm dim~ \rm H_{abs}^{(0)}(p_{n,n'}) = 1. \eqno (3.3) $$
Call a representative of the non-trivial cohomology class $\vert \sl W_{\rm
n,n'}\rangle$.  We will now show that
$$ \Omega _{\rm n,n'} =  \sl W_{\rm n,n'} \sl \bar O_{\rm n,n'} + \sl
O_{\rm n,n'} \bar W_{\rm n,n'} \eqno (3.4)$$
is the zero-form component of a current, in the formalism described in section
2.2.  In order to do this, we must prove that $\vert \Omega _{\rm n,n'}\rangle$
is in the
semi-relative BRST cohomology and that $b_{-1}\vert \Omega _{\rm n,n'}\rangle
\not= \rm 0$.

Consider the state $b_{0}\vert \sl W_{\rm n,n'} \rangle$.  We know this does
not
vanish, because there is no relative cohomology with momentum $p_{n,n'}$
at ghost number 0.  However, $Q(b_{0}\vert \sl W_{\rm n,n'} \rangle)
= \rm 0$, since $[Q,b_{0}] = L_{0}$ and $\vert \sl W_{\rm n,n'} \rangle$ is
annihilated by both $Q$ and $L_{0}$. In addition, $b_{0} \vert \sl
W_{\rm n,n'} \rangle$ is obviously annihilated by $b_{0}$, so it is in
the relative cohomology at ghost number -1.  Since $\vert \sl O_{\rm
n,n'} \rangle$ is the only nontrivial relative cohomology at ghost number
-1
with
the right momentum, we must have:
$$ b_{0} \vert \sl W_{\rm n,n'}\rangle = \rm k \vert \sl O_{\rm
n,n'}\rangle,~\rm k
\not= 0. \eqno (3.5)$$
But then consider the state in the full closed string cohomology given
by:
$$\vert \Omega _{\rm n,n'} \rangle = \vert \sl O_{\rm n,n'} \rangle
\otimes \vert \sl \bar W_{\rm n,n'} \rangle + \vert \sl W_{\rm
n,n'} \rangle \otimes \vert \sl \bar O_{\rm n,n'}\rangle~.\eqno (3.6)$$

This is clearly annihilated by $Q_{TOT}$ and $b_{0} - \bar b_{0}$,
so it is in the semi-relative cohomology of the closed string theory.
Hence, the operator $\Omega _{\rm n,n'}$ given in (3.4) corresponding
to this state is the zero form component of a current if $b_{-1}\vert
 \Omega _{\rm n,n'} \rangle \not= \rm 0$.  We shall prove that this is
the case in section 3.2, by making a connection with previous work in
the c=1 theory which allows us to give a more explicit construction of
these states.
\subsec{Comparison to c=1 Model}
We might expect these currents to correspond in a simple way to some
subalgebra of the
infinite symmetry algebra present in the c=1 theory.  This is in fact
the case.

Recall that in the c=1 theory, the symmetries include a $W_{\infty}$
algebra but also include an infinite number of `internal' symmetries,
one for each ground ring element. If we define following \Li
$$\sl a~=~[Q,\phi] = c\partial \phi~+~{\sqrt 2} \partial c\eqno
(3.7)$$
then the new chiral ghost number one operators in the absolute
cohomology of the c=1 model were given by
$$\sl a O(0) = {1 \over 2\pi i} \oint {dz\over z} a(z)~O(0)\eqno
(3.8)$$
where $\sl O$ is an element of the c=1 chiral ground ring \Zwi.  While it might
seem
that these operators are BRST trivial from (3.7), this is not so
strictly speaking because $\phi$ is not in the usual space of conformal
fields.  The new symmetry generators then had zero form pieces corresponding to
the new
semi-relative cohomology classes $\sl (a~+~\bar a)(O\bar O)$.

Because the Liouville sector remains intact in the minimal models
coupled to gravity, one can still construct the operator $\sl a$ and
this construction of the new semi-relative cohomology should generalize to all
of the minimal models.
Hence, we find that our new operators $\Omega _{n,n'}$ are given
more explicitly by the formula
$$\Omega _{\rm n,n'}~=~\sl (a + \bar a)R_{\rm n,n'}.\eqno (3.9)$$
This allows us to fix the constant $k$ in equation (3.5) to be $\sqrt
2$.
It is also now a trivial matter to check the remaining condition which
insures that $\Omega _{\rm n,n'}$ generates a non-trivial symmetry,
namely that $b_{-1}\vert \Omega _{\rm n,n'}\rangle \not= \rm 0$.  From
equations (3.7), (3.9)
we see with a little bit of work that the symmetries are in fact nontrivial
(in this sense, at least).

Although we have exhibited that the ($p,p^\prime$) model still has a Lie
algebra
of symmetries with $(p-1)(p^\prime-1)$ generators, one may wonder what has
become of the rest of the vast symmetry structure of the c=1 theory.
Indeed, if one views the ($p,p^\prime$) models as  SO(2,$\cal C$) rotations of
the c=1
model, one might expect the $W_{\infty}$ structure to survive \CDK \Dot.
Nevertheless, it follows from the mathematical analysis \BMP ~that in the
($p,p^\prime$) model there are
no potential generators of conventional (vanishing ghost number) symmetries
other than those given above.
\subsec{Generalization to Other Ghost Numbers}
It is clear that the construction of sections 3.1 and 3.2 will generalize to
other ghost numbers as follows.  Assume we have some momentum p and ghost
number
n such that:
$$\rm dim~ H_{rel}^{(n)}(p)~=~1.\eqno (3.10)$$
Suppose a representative of this cohomology class is $\vert A_{\rm
n}^{p}\rangle$.  Then as discussed in section 3.2, we can construct a new
element of the semi-relative cohomology at ghost number n+1:
$$(\sl a + \sl \bar a)A_{\rm n}^{\rm p}\bar A_{\rm n}^{\rm p}(0)\vert 0
\rangle.\eqno (3.11)$$
These states will correspond to `symmetries' of non-standard ghost
number. The associated conserved charges will map `normal' physical
states to physical states of exotic ghost number.
\newsec{Algebra of the Symmetries}
\subsec{Role of New Symmetries in D=2 Case}
Before proceeding to determine the symmetry structure of the $D<2$
theories, it will be helpful to recall certain facts revealed in the
analysis of \Zwi~ about the analogues of our new symmetry generators in
the D=2 case.

In the (compactified) c=1 model, the chiral ground ring is generated by two
operators
$\sl x$ and $\sl y$, while the anti-chiral ground ring is generated by
their counterparts $\sl x'$ and $\sl y'$.  We can combine these chiral
components to form four closed-string operators:
$$\sl a_{\rm 1} = \sl xx'~~~a_{\rm 2}=\sl yy'~~~a_{\rm 3}=\sl xy'~~~a_{\rm
4}=yx' . \eqno (4.1)$$
 In \GR \Zwi~ the special
states are interpreted in terms of the differential geometry of the quadric
cone
$\cal Q$ formed by the ring generators subject to the `fermi-surface' relation
$$\sl a_{\rm 1}a_{\rm 2}~-~a_{\rm 3}a_{\rm 4} = \mu \eqno (4.2)$$
where $\mu$ is the cosmological constant in the Liouville theory.
In this context, the new symmetry currents derived from the zero form $(a +
\bar a)O$
(with $\sl O$ a ground ring element) correspond to vector fields
$${\rm Currents~ from~} (\sl a +\sl \bar a)O \sim f~S,~~~~S\equiv x{\partial
\over {\partial x}} +
y{\partial \over {\partial y}} - x'{\partial \over {\partial x'}} -
y'{\partial \over {\partial y'}}~.\eqno (4.3)$$
Here, $f$ is some function on $\cal Q$ which must have the
proper quantum numbers to represent the symmetry generator in question.

We will use this information about the D=2 theory to help us determine
the
algebra of the new symmetries in $D<2.$
\subsec{Symmetry Algebra in $D<2$}
The dimension 1, ghost number zero currents $\cal J_{\rm n,n'}$ in the closed
string theory
(the one-form components in the terminology of section 2.2) are defined
using the descent equations applied to the zero forms $\Omega$ given in
equation (3.9):
$$\Omega_{\rm n,n'} \rightarrow \cal J_{\rm n,n'}~ {\rm via~ descent
{}~equations.}\eqno (4.4)$$
We know from the explicit construction of the ground ring generators
\KMS~ that
$$ p_{n,n'} = - [ (n-1) + (p/p') (n'-1) ] {\gamma \over 2}~~~~~\gamma =
\sqrt {2p' \over p} \eqno
(4.5)$$
is the Liouville momentum of $\cal J_{\rm n,n'}$.  Hence, by momentum
conservation alone it is clear that the symmetry algebra must be of the
form
$$
[\cal J_{\rm m,n},~\cal J_{\rm m',n'}]~=\rm~F_{\rm
m,n~m',n'}^{m+m'-1,n+n'-1}~\cal
J_{\rm m+m'-1,n+n'-1}\eqno
(4.6)$$
where it is understood that $\cal J_{\rm a,b} \equiv \rm 0$ unless $1\le
a\le p-1$ and
$1\le b \le p^\prime -1$.
Since explicit forms of these operators are not available (however see \imbimbo
\govind ~for recent progress in constructing explicit representatives of
BRST cohomology classes for $c<1$ matter coupled to gravity), we must
find some means other than direct computation of determining the
non-vanishing
structure constants.  This is where the interpretation of section 4.1
will be useful.

Just as in the D=2 theory, we can think of the ring in the $D<2$ case as
having four chiral generators in terms of which $\sl X$ and $\sl Y$ are
defined:
$$ \sl X = R_{\rm 2,1} = xx'~~~~~~~\sl Y = R_{\rm 1,2} = yy' .\eqno (4.7)$$
Of course, in view of the conditions in the ($p,p^\prime$) model that $\sl
X^{\rm p-1}=Y^{\rm p'-1} =\rm 0$
we should also set
$$x^{\rm p-1} = x'^{\rm p-1} = y^{\rm p'-1} =
y'^{\rm p'-1} = 0.\eqno (4.8)$$

Now, by analogy with the D=2 theory we expect the new symmetries should
act on the states of the $D<2$ models as polynomials in
the new ring generators multiplied by the analogue of S (defined in
equation (4.3)).  Thus, we make the identification
$$\cal J_{\rm n,n'} = \sl f_{\rm n,n'}~S\eqno (4.9)$$
with the understanding that the variables
$\sl x,x',y,y'$ in the definition of S are now to be interpreted as defined
above for the ($p,p^\prime$)
model.  Here, $f_{\rm n,n'}$ is a function with the same quantum numbers
as the operator on the left hand side.  As it is a polynomial function of the
ring generators, we gather that f must have the form
$$\sl f_{\rm n,n'} \sim X^{\rm n-1}Y^{\rm n'-1} = (xx')^{\rm
n-1}(yy')^{\rm n'-1}.\eqno (4.10)$$

Hence, up to normalization we can find the current algebra by taking the
commutators of the associated vector fields
$$[\cal J_{\rm n,n'}, \cal J_{\rm m,m'}] = \sl [f_{\rm n,n'}S,f_{\rm
m,m'}S]\eqno (4.11)$$
with the result that
$$[\cal J_{\rm n,n'}, \cal J_{\rm m,m'}] \sim \rm (n+n'-m-m') \cal J_{\rm
n+m-1, n'+m'-1}\eqno (4.12)$$
with the same understanding as after equation (4.6) about the indices on
the currents.
\subsec{Conjectured Algebra of Exotic Symmetries}
As discussed in section 3.3, our construction of closed string symmetries will
generalize to all cases where there is isolated chiral relative
cohomology at a given ghost number and momentum. It is known from the
mathematical analysis \LZ \BMP ~ that there is in fact $(p-1)(p^\prime
-1)$ dimensional relative cohomology at all negative ghost numbers in
the ($p,p^\prime)$ model.   If we denote the dimension one currents
obtained from this cohomology at ghost number k as $\cal J^{\rm k}_{\rm
n,n'}$, with $1\le n \le p-1, 1\le n^\prime \le p^\prime -1$, then the
natural conjecture for the extended symmetry algebra is simply
$$[\cal J^{\rm k}_{\rm n,n'}, \cal J^{\rm j}_{\rm m,m'}] \sim \rm
(n+n^\prime - m - m^\prime)\cal J^{\rm j+k}_{\rm n+m-1, n'+m'-1}
\eqno (4.13)$$
with $j,k \le 0$, and again with the implicit understanding that $\cal J
\equiv \rm 0$ if its indices do not lie in the standard allowed range for
the ($p,p^\prime$) model.
\newsec{Action of Conventional Charges on States}
\subsec{States and Currents in $D<2$}
It is obvious from the form of S, equation (4.3), that the most conventional
symmetries
we have constructed (those with vanishing total ghost number) annihilate the
ring states.  In fact, their associated charges also annihilate the normal
`tachyon'  states obtained by dressing the primary fields of the minimal
models.

For the ($p,p^\prime$) model, the spectrum of tachyon states is
$$\cal T_{\rm n,n'} = \rm c\bar c \cal O_{\rm n,n'}\rm e^{[{1 + {p \over
p'} - {{pn' - p'n}\over p'}]{\gamma \over 2}\phi}}\eqno (5.1)$$
(with $1\le n \le p-1,~1\le n^\prime \le p^\prime -1$) where $\cal O_{\rm
n,n'}$ is a matter primary field with $\cal O_{\rm
n,n'} = \cal O_{\rm p-n, p'-n'}$ (so we identify
$\cal T_{\rm n,n'} = \cal T_{\rm p-n, p'-n'}$).  Of course, there are
also $(p-1)(p^\prime -1)$ special states at every negative value of the ghost
number for which we have no such convenient representation.

To actually construct the charges $$ Q_{\rm n,n'} = {1\over 2\pi i}~\oint \cal
J_{\rm
n,n'}\eqno (5.2)$$  associated with $\cal J_{\rm n,n'}$,
we first use the descent equations to get a more explicit form of $\cal
J_{\rm n,n'}$.  If we define:
$$\partial O_{\rm n,n'} = \{ Q_{TOT},Z_{n,n'}\},~~~\partial (\sl aO_{\rm
n,n'})=\{Q_{\rm TOT},Y_{\rm
n,n'}\}\eqno (5.3)$$
then the descent equations tell us that
$$\cal J_{\rm n,n'} = \sl (Y_{\rm n,n'}{\bar O_{\rm n,n'}} + Z_{\rm
n,n'}{\sl \bar a \bar O_{\rm n,n'}}) dz + (\sl O_{\rm n,n'} \bar Y_{\rm n,n'} -
\sl aO_{\rm n,n'}\bar Z_{\rm n,n'}) d\bar z.\eqno (5.4)$$

Using this form of $\cal J_{\rm n,n'}$, we will argue that $\sl Q_{\rm
n,n'}$ must annihilate the tachyon states (5.1).
\subsec{Action of Conventional Charges on Tachyon States}
{}From (5.4) it follows immediately that
$$Q_{\rm n,n'} = \oint {dz \over 2\pi i}( Y_{\rm n,n'}\bar O_{\rm n,n'} +
Z_{\rm n,n'}{\sl \bar a \bar O_{\rm n,n'}})~+~\oint {d\bar z \over 2\pi i}
(O_{\rm n,n'}\bar Y_{\rm n,n'} - \sl aO_{\rm n,n'}\bar Z_{\rm n,n'})\eqno
(5.5)$$
are the conserved charges.

Notice that although $Q_{\rm n,n'}$ does indeed have vanishing total ghost
number, this arises in a peculiar way: The terms containing $Y$
and $O$ have left-right ghost numbers $(0,0)$, but the terms containing $Z$
and $\sl aO$ have left-right ghost numbers $(-1,1)$ and $(1,-1)$. While in the
D=2 theory these types of operators can map special tachyons to the new
states which contain parts of ghost number $(1,-1)$ and $(-1,1)$ (such
states do exist in the D=2 theory, see \Zwi),
there are no such states in $D<2$.
Hence, it
follows immediately that the pieces in $Q_{\rm n,n'}$ of this type act
trivially on the tachyon states in $D<2$.  So we need only consider the
action of the terms containing $Y\bar O$ and $O \bar Y$.

However, the result of applying $Q_{\rm n,n'}$ to a tachyon state must
be another tachyon state (there are no other candidates in $D<2$) and
hence must be left-right symmetric.  Recalling our conventions for
contour integrals (see section 2.2), it is evident that any left-right
symmetric pieces which emerge from applying the $Y\bar O$ and $O \bar Y$
terms to a tachyon state will exactly cancel each other.  Hence, the
charges $Q_{\rm n,n'}$ annihilate the tachyon states (5.1).
\subsec {Are the Charges Physically Non-trivial?}
Since we have seen that the conventional charges annihilate the ring states
$\it and$
the tachyon states, one might wonder if they are physically relevant.
At any rate, it seems unlikely that the full extended symmetry algebra
discussed in section
4.3 will act trivially on the entire tower of
special
states obtained in \LZ \BMP, but we have no convincing
evidence regarding this matter.  If it does, one would have essentially
another infinite tower of states (corresponding to semi-relative
cohomology classes) in the $D<2$ closed string theories
which are effectively decoupled from the conventional states in the
relative cohomology.

\newsec{Conclusion}
There are several striking features of the $c<1$ matrix models which
we would like to understand from the continuum perspective.  In
particular, the fact that the square root of the partition function $\cal Z$
is a $\tau$ function of the KdV hierarchy \Douglas, and the emergence of an
infinite number of Virasoro and W-constraints which also uniquely
determine $\cal Z$ \DVV, are still mysterious from the perspective of
Liouville theory coupled to minimal matter (however see \KMS$~$  for some
interesting conjectures relating the ground ring to Douglas' operators
P,Q \Douglas).

It seems plausible that the incorporation of all of the `exotic'
currents found here into an extended symmetry algebra, as discussed
briefly in section 4.3, might shed some light on these issues.
However, the role of these symmetries in the $D<2$ theory is not yet
well understood.  We have argued that the most conventional ones
act trivially on the ring and tachyon states, but the interesting question of
the action of the full symmetry algebra on the entire tower of states in
the ($p,p^\prime$) model remains open.

To extract a deeper understanding of string theory from the matrix models, it
is important that we reproduce as much of their structure from the
continuum Liouville $\times$ Matter viewpoint as we can.  Uncovering the
KdV/Virasoro structure in the continuum formulation will help us to
understand the `miracles' of the matrix models in terms of conventional
string theory, and is certainly a problem which warrants
further study.

$$\underline {\bf Acknowledgements}$$

I would like to thank J. Distler for
introducing me to several issues related to the subject of this
paper, and M. Doyle for correcting my understanding of the operator
formalism. I am also grateful to I. Klebanov for helpful
discussions and several very useful remarks.
I owe  special thanks to E. Witten for encouraging me to
investigate this topic
and for stimulating discussions.

\listrefs
\end